\documentclass[preprint,amsmath,amssymb,aps,prd,floatfix,superscriptaddress,nofootinbib]{revtex4-1}

\usepackage{colortbl}
\usepackage{graphicx}
\usepackage{makecell}
\usepackage{mathtools}
\usepackage{multirow}
\usepackage{makecell}
\usepackage{mathtools}
\usepackage{appendix}
\usepackage{booktabs}
\usepackage{physics}
\usepackage[colorlinks,linkcolor=blue,anchorcolor=blue,citecolor=blue, urlcolor=blue]{hyperref}

\begin{document}
\title{Testing $n_s=1$ in light of the latest ACT and SPT data}

\author{Ze-Yu Peng}
\email{pengzeyu23@mails.ucas.ac.cn}
\affiliation{School of Physical Sciences, University of Chinese Academy of Sciences, Beijing 100049, China}
\affiliation{International Centre for Theoretical Physics Asia-Pacific, University of Chinese Academy of Sciences, 100190 Beijing, China}

\author{Jun-Qian Jiang}
\email{jiangjq2000@gmail.com} \affiliation{School of Physical
Sciences, University of Chinese Academy of Sciences, Beijing
100049, China}

\author{Hao Wang}
\email{wanghao187@mails.ucas.ac.cn} \affiliation{School of
Physical Sciences, University of Chinese Academy of Sciences,
Beijing 100049, China}

\author{Yun-Song Piao}
\email{yspiao@ucas.ac.cn}
\affiliation{School of Physical Sciences, University of Chinese Academy of Sciences, Beijing 100049, China}
\affiliation{International Centre for Theoretical Physics Asia-Pacific, University of Chinese Academy of Sciences, 100190 Beijing, China}
\affiliation{School of Fundamental Physics and Mathematical Sciences, Hangzhou Institute for Advanced Study, UCAS, Hangzhou 310024, China}
\affiliation{Institute of Theoretical Physics, Chinese Academy of Sciences, P.O. Box 2735, Beijing 100190, China}

\begin{abstract}

It is commonly recognized that the primordial scalar spectral index $n_s$ is approximately $0.96-0.975$, depending on the dataset. However, this view is being completely altered by the early dark energy (EDE) resolutions of the Hubble tension, known as the most prominent tension the standard $\Lambda$CDM model is suffering from.
In corresponding models with pre-recombination EDE, resolving the Hubble tension (i.e., achieving $H_0\sim 73$km/s/Mpc) must be accompanied by a shift of $n_s$ towards unity to maintain consistency with the cosmological data, which thus implies a scale invariant Harrison-Zel'dovich spectrum with $n_s=1$ $(|n_s-1|\simeq {\cal O}(0.001))$. In this work, we strengthen and reconfirm this result with the latest ground-based CMB data from ACT DR6 and SPT-3G D1, the precise measurements at high multipoles beyond the Planck angular resolution and sensitivity. Our work again highlights the importance of re-examining our understanding on the very early Universe within the broader context of cosmological tensions.

\end{abstract}

\maketitle

\newpage

\section{Introduction}\label{sec:intro}

The spectral index of primordial scalar perturbations, $n_s$, is
the most crucial parameter for understanding the physics of
inflation. The Planck collaboration using their cosmic microwave
background (CMB) data has precisely constrained its value to $n_s
= 0.965\pm 0.004$ (68\% CL) \cite{Planck:2018vyg}, and ruled out
the scale-invariant Harrison-Zel'dovich (HZ) spectrum ($n_s=1$) at
more than $8\sigma$ significance level.

However, this seemingly conclusive result is based on the standard
$\Lambda$CDM model, which is currently suffering from
observational tensions. The most prominent among them is the
Hubble tension
\cite{Verde:2019ivm,Perivolaropoulos:2021jda,DiValentino:2021izs,Schoneberg:2021qvd,Shah:2021onj,Abdalla:2022yfr,DiValentino:2022fjm,Verde:2023lmm},
%which refers to the significant discrepancy between the Hubble
%constant ($H_0$) inferred from early-Universe probes like the CMB,
%and the value determined from local measurements, such as those
%from Cepheid-calibrated type Ia supernovae (SN)
%\cite{Riess:2021jrx}. The persistence of this discrepancy
which has led to a consensus that new physics beyond $\Lambda$CDM
might be required
\cite{Mortsell:2018mfj,Vagnozzi:2019ezj,Knox:2019rjx,Hu:2023jqc}.
A compelling resolution of the Hubble tension is Early Dark Energy
(EDE)
\cite{Poulin:2018cxd,Kaloper:2019lpl,Agrawal:2019lmo,Lin:2019qug,Smith:2019ihp,Niedermann:2019olb,Sakstein:2019fmf,Ye:2020btb,Gogoi:2020qif,Braglia:2020bym,Lin:2020jcb,Odintsov:2020qzd,Seto:2021xua,Ye:2021iwa,Nojiri:2021dze,Karwal:2021vpk,Wang:2022jpo,Rezazadeh:2022lsf,Poulin:2023lkg,Sohail:2024oki}.
In corresponding EDE models, an energy component is non-negligible
only for a short epoch before recombination, which suppressed the
comoving sound horizon at recombination, and thus makes the CMB
and baryon acoustic oscillations (BAO) data reconciled with a high
Hubble constant $H_0\gtrsim 70$km/s/Mpc. In particular, AdS-EDE
\cite{Ye:2020btb}, which incorporates an anti-de Sitter (AdS)
phase around recombination, can lead to $H_0\sim 73$km/s/Mpc,
since it allows a more efficient injection of EDE
\cite{Ye:2020oix,Jiang:2021bab,Ye:2022efx}.

It is usually thought that new physics beyond $\Lambda$CDM did not
have a significant impact on $n_s$ ($n_s\simeq 0.96-0.975$
dependent of different CMB and BAO datasets), however, the
injection of EDE before the recombination completely altered this
cognition. It has been found that in corresponding scenario $n_s$
positively correlates with $H_0$, and scales as \cite{Ye:2021nej}:
\begin{equation} \label{eq:deltans}
    {\delta n_s}\simeq 0.4\frac{\delta H_0}{H_0},
\end{equation}
which suggests that $n_s$ must significantly shift towards $n_s=1$
in such $\Lambda$CDM+EDE models\footnote{The possibilities of
$n_s=1$ in different cases have been also investigated in
Refs.\cite{DiValentino:2018zjj,Giare:2022rvg,Calderon:2023obf}.}.
As a result, complete EDE solutions of the Hubble tension seem to
be pointing to a scale-invariant HZ spectrum, i.e.$n_s= 1$, for
$H_0\simeq 73$km/s/Mpc
\cite{Ye:2020btb,Jiang:2022uyg,Jiang:2022qlj,Wang:2024dka,Wang:2024tjd}.
This finding is also consistent with Planck-independent CMB data
\cite{Jiang:2022uyg,Smith:2022hwi,Peng:2023bik}, including earlier
Atacama Cosmology Telescope (ACT) \cite{ACT:2020gnv} and South
Pole Telescope (SPT) \cite{SPT-3G:2021eoc,SPT-3G:2022hvq} data.
See also \cite{Forconi:2025zzu} for a recent work that analyzes different extensions with the HZ spectrum.

Recently, both ACT and SPT have released their new data
\cite{ACT:2025fju,ACT:2025tim,SPT-3G:2025bzu}, which are the most
precise measurements of small-scale CMB polarization to date.
Their combination with Planck data yields the tightest CMB
constraints, showing no evidence for physics beyond $\Lambda$CDM.
It is therefore timely and crucial to revisit the scale relation
(\ref{eq:deltans}) and the implications of EDE models for $n_s$ in
light of latest ACT and SPT data, see
\cite{Poulin:2025nfb,SPT-3G:2025vyw} for recent works on
axion-like EDE.

In this work, we test whether $n_s=1$ for $H_0\simeq 73$km/s/Mpc
is still robust with the latest ACT and SPT data. We consider two
representative EDE models, axion-like EDE and AdS-EDE. The rest of
the paper is organized as follows: In Sec.~\ref{sec:ns}, we review
the $n_s-H_0$ scaling relation and its prediction for $n_s=1$. We
present our results in Sec.~\ref{sec:results}, including the
datasets and methods used and the constraints on axion-like EDE
and AdS-EDE. Finally, we discuss the implications of our findings
and conclude in Sec.~\ref{sec:discussion}.

\section{$n_s= 1$}\label{sec:ns}

It is necessary to reclarify why the scaling relation
(\ref{eq:deltans}) exists in pre-recombination resolutions of the
Hubble tension, since it straightly implies $n_s= 1$.

%For example, a higher value of $\omega_{\mathrm{cdm}}$ is needed
%to offset the enhanced early Integrated Sachs-Wolfe (ISW) effect
%caused by EDE, which in turn leads to a higher
%$S_8$~\cite{Vagnozzi:2021gjh}.

The damping angular scale
\begin{equation}
\theta_D^*=\frac{r_D^*}{D_A^*} \sim r_D^*H_0,
\end{equation}
where $r_D^*$ is the damping scale at recombination and $D_A^*\sim
1/H_0$ is the angular diameter distance to the last scattering
surface, has been precisely measured by the CMB. Thus to make
$H_0$ higher but not spoil the fit to CMB, a smaller $r_D^*$, just
like the sound horizon, is required. It is known that the damping
scale at recombination is $r_D^*\sim
\omega_b^{-1/2}\omega_{\mathrm{cdm}}^{-1/4}$
\cite{2020moco.book.....D}, thus we have
\begin{equation} \theta_D^* \sim
\omega_b^{-1/2}\omega_{\mathrm{cdm}}^{-1/4}H_0.
\end{equation}
In fact, $\Omega_{\mathrm{cdm}}=\omega_{\mathrm{cdm}}H_0^{-2}$ is
well constrained by CMB and BAO data, which implies
$\omega_b^{-1}H_0\simeq\mathrm{const}$, thereby requiring a higher
baryon density $\omega_b$ for a higher $H_0$. This higher
$\omega_b$ enhances the baryon loading effect, magnifying the
ratio between the first and second acoustic peak of the CMB TT
spectrum, which must be compensated by a larger spectral index,
with $\delta n_s \simeq 0.8 \delta\omega_b/\omega_b$.
Consequently, Ref.~\cite{Ye:2021nej} unveiled an universal
$n_s-H_0$ scaling relation:
\begin{equation}
    \delta n_s \simeq 0.8(1-\alpha) \frac{\delta H_0}{H_0}
\end{equation}
where $\alpha$ parameterizes the additional damping needed to
accommodate a larger $n_s$\footnote{Any pre-recombination solution
to the Hubble tension that suppressed the sound horizon, including
EDE, inevitably requires compensatory shifts in other cosmological
parameters. See also Ref.~\cite{Jiang:2024nha} for a summary of
the reasons behind the shift of $n_s$.}.

Specifically, for the Planck+BAO+Pantheon dataset, the spectral
index $n_s$ scales as in Eq.~\eqref{eq:deltans} ($\alpha\simeq
0.5$) \cite{Ye:2021nej}, while for Planck+(earlier
ACT+SPT)+BAO+Pantheon dataset, it scales as:
\begin{equation}\label{ACTSPTns}
    \delta n_s \simeq 0.3 \frac{\delta H_0}{H_0},
\end{equation}
with a slightly smaller scale factor
\cite{Jiang:2022uyg,Smith:2022hwi,Peng:2023bik,Toda:2025kcq}. As a result, a
Hubble constant around $H_0\simeq 73$ km/s/Mpc would correspond to
a scale-invariant HZ spectrum ($n_s= 1$).

\section{Testing $n_s=1$ in light of latest data}\label{sec:results}

\subsection{Datasets and Methods}\label{sec:methods}
% Inspired by \cite{SPT-3G:2025bzu}, we combine the ground-based data from ACT DR6\footnote{The \texttt{ACT-lite} likelihood from \url{https://github.com/ACTCollaboration/DR6-ACT-lite}} \cite{ACT:2025fju, ACT:2025tim} and SPT-3G D1\footnote{The \texttt{SPT-lite} likelihood from \url{https://github.com/SouthPoleTelescope/spt_candl_data}} \cite{SPT-3G:2025bzu,Balkenhol:2024sbv} with the large-scale Planck 2018 CMB data\footnote{The \texttt{Plik-lite} likelihood from \url{https://github.com/ACTCollaboration/DR6-ACT-lite}} \cite{Planck:2019nip} truncated at $\ell<1000$ in TT, and $\ell<600$ in TE and EE, denoted as \textbf{Planck+SPT+ACT}. This dataset also include the CMB lensing data\footnote{The
% \texttt{actplanckspt3g\_baseline} variant of
% \url{https://github.com/qujia7/spt_act_likelihood}} from Planck PR4  \cite{Carron:2022eyg}, ACT DR6 \cite{ACT:2023kun,ACT:2023dou,ACT:2023ubw} and SPT-3G \cite{SPT-3G:2024atg,SPT-3G:2025zuh}.
% For comparison, we also consider the full Planck  data, including the CMB lensing data from Planck PR4, denoted as \textbf{Planck}.
% For both cases, we include the Planck \texttt{Commander} likelihood for low-$\ell$ TT spectrum, and the \texttt{SimALL} likelihood for low-$\ell$ EE spectrum \cite{Planck:2019nip}.

Inspired by \cite{SPT-3G:2025bzu}, we combine the ground-based ACT
DR6 \cite{ACT:2025fju, ACT:2025tim} and SPT-3G D1
\cite{SPT-3G:2025bzu,Balkenhol:2024sbv} data with the large-scale
Planck 2018 data \cite{Planck:2019nip}, which is denoted as
\textbf{Planck+SPT+ACT}. We also consider the full Planck data,
denoted as \textbf{Planck}, for comparison. The details of both
CMB datasets used are presented in Table~\ref{tab:datasets}.

\begin{table}[htbp]
    \centering
    \begin{tabular}{|l|p{0.75\linewidth}|}
        \hline
        \textbf{Dataset} & \textbf{Description} \\
        \hline
\textbf{Planck} & The CMB-only \texttt{Plik-lite} likelihood for Planck 2018 high-$\ell$ TT/TE/EE spectra\cite{Planck:2019nip} + Planck \texttt{Commander} and \texttt{SimALL} likelihood for low-$\ell$ TT and EE spectra \cite{Planck:2019nip} + CMB lensing data from Planck PR4 \cite{Carron:2022eyg} \\
        \hline
\textbf{Planck+SPT+ACT} & \texttt{ACT-lite} likelihood for ACT DR6 \cite{ACT:2025fju, ACT:2025tim} + \texttt{SPT-lite} likelihood for SPT 3G D1 \cite{SPT-3G:2025bzu,Balkenhol:2024sbv} + \texttt{Plik-lite} likelihood cut at $\ell>1000$ in TT, and $\ell>600$ in TE and EE + Planck \texttt{Commander} and \texttt{SimALL} likelihood for low-$\ell$ TT and EE spectra \cite{Planck:2019nip} + CMB lensing data from Planck PR4  \cite{Carron:2022eyg}, ACT DR6 \cite{ACT:2023kun,ACT:2023dou,ACT:2023ubw} and SPT-3G \cite{SPT-3G:2024atg,SPT-3G:2025zuh}. \\
        \hline
    \end{tabular}
    \caption{The CMB datasets used in this work. Both datasets also include DESI BAO data and Pantheon+ SN data with and without SH0ES calibration.}
    \label{tab:datasets}
\end{table}

Both datasets also include the \textbf{DESI} DR2 BAO data
\cite{DESI:2025zgx}. In addition, we consider the uncalibrated
Type Ia SN from the \textbf{Pantheon+} dataset
\cite{Scolnic:2021amr}, which is compared to the SH0ES Cepheid
calibrated dataset, \textbf{Pantheon+SH0ES} \cite{Riess:2021jrx}.

To test the $n_s-H_0$ scaling relation \eqref{eq:deltans}, in
particular $n_s=1$ for $H_0\simeq 73$km/s/Mpc, we focus on the EDE
models. Besides the original axion-like EDE model
\cite{Poulin:2018dzj,Poulin:2018cxd}, we also consider the AdS-EDE
model
%which is known for yielding a large $H_0$ even without any
%late-time $H_0$ prior
\cite{Ye:2020btb,Ye:2020oix,Jiang:2021bab,Ye:2022efx}. The details
of both models are presented in Appendix~\ref{sec:ede}. We perform
the Markov chain Monte Carlo (MCMC) analysis using \texttt{Cobaya}
\cite{Torrado:2020dgo}. The observables are computed using the
cosmological Boltzmann code \texttt{CLASS} \cite{Blas:2011rf} and its modified version\footnote{We use AxiCLASS
(\url{https://github.com/PoulinV/AxiCLASS}) for axion-like EDE and classmultiscf
(\url{https://github.com/genye00/class_multiscf.git}) for AdS-EDE.}. We
adopt wide, flat priors for all relevant parameters, as presented
in Table~\ref{tab:priors}. We take our MCMC chains to be converged
using the Gelman-Rubin criterion \cite{Gelman:1992zz} with
$R-1<0.05$.

\begin{table}[htbp]
    \centering
    \begin{tabular}{|l|c|}
        \hline
        Parameter                & Prior                \\
        \hline
        $f_\mathrm{EDE}(z_c)$        & $[0,\, 0.5]$          \\
        $\log_{10}(z_c)$             & $[3,\, 4]$          \\
        $\theta_\mathrm{ini}$        & $[0,\, 3.1]$          \\
        \hline
        $\log(10^{10} A_\mathrm{s})$ & $[1.61,\, 3.91]$  \\
        $n_\mathrm{s}$           & $[0.8,\, 1.2]$       \\
        $H_0$                    & $[20,\, 100]$        \\
        $\Omega_\mathrm{b} h^2$  & $[0.005,\, 0.1]$     \\
        $\Omega_\mathrm{c} h^2$  & $[0.001,\, 0.99]$    \\
        $\tau_\mathrm{reio}$     & $[0.01,\, 0.8]$      \\
        \hline
    \end{tabular}
\caption{The priors for relevant parameters in our MCMC analysis.
For both EDE models, $z_c$ is the critical redshift at which EDE
starts to decay and $f_\mathrm{EDE}(z_c)$ is the fraction of EDE
energy density at $z_c$. In addition, $\theta_\mathrm{ini}$ is the
initial value of the EDE field in axion-like EDE, and following
\cite{Ye:2020btb} we fix the depth of the AdS well to
$\alpha_{\mathrm{AdS}}
\equiv\left(\rho_{\mathrm{m}}\left(z_c\right)+\rho_{\mathrm{r}}\left(z_c\right)\right)
V_{\mathrm{AdS}}=3.79 \times 10^{-4}$ in AdS-EDE.}
    \label{tab:priors}
\end{table}

%\subsection{Axion-like EDE}

\subsection{Result for both axion-like and AdS EDEs}

The mean and $1\sigma$ errors of cosmological parameters are
presented in Tables~\ref{tab:axion} and \ref{tab:ads} f.                                         or
axion-like EDE and AdS-EDE, respectively.

\begin{table}[htbp]
    \centering
    \begin{tabular}{|l|c|c|c|c|}
        \hline
        \multicolumn{1}{|l|}{Parameter} & \multicolumn{2}{c|}{Planck} & \multicolumn{2}{c|}{Planck+SPT+ACT} \\
        \cline{2-5}
        & w/o SH0ES & w/ SH0ES & w/o SH0ES & w/ SH0ES \\
        \hline
        $f_\mathrm{EDE}(z_c)$        & $<0.107$                  & $0.127\pm0.024$           & $<0.105$                  & $0.120\pm0.020$           \\
        $\log_{10}(z_c)$             & $3.60^{+0.23}_{-0.19}$    & $3.611^{+0.013}_{-0.088}$ & $3.50\pm0.15$             & $3.548^{+0.032}_{-0.038}$ \\
        $\theta_\mathrm{ini}$        & ---                       & $2.75^{+0.12}_{-0.076}$   & ---        & $2.73^{+0.10}_{-0.075}$         \\
        \hline
        $H_0$                        & $69.58^{+0.61}_{-1.3}$    & $72.29\pm 0.82$   & $69.50^{+0.70}_{-1.2}$    & $71.95\pm 0.71$            \\
        $100\Omega_\mathrm{b} h^2$   & $2.266^{+0.017}_{-0.021}$ & $2.285\pm0.022$           & $2.255\pm0.013$           & $2.268\pm0.013$ \\
        $\Omega_\mathrm{c} h^2$      & $0.1223^{+0.0018}_{-0.0045}$ & $0.1315\pm 0.0033$       & $0.1229^{+0.0024}_{-0.0041}$ & $0.1308\pm 0.00277$     \\
        $10^9A_\mathrm{s}$           & $2.127\pm0.031$           & $2.155\pm 0.031$           & $2.145\pm0.026$           & $2.160^{+0.024}_{-0.026}$           \\
        $n_\mathrm{s}$               & $0.9774^{+0.0053}_{-0.0086}$ & $0.9921^{+0.0057}_{-0.0064}$      & $0.9792^{+0.0050}_{-0.0060}$ & $0.9897^{+0.0045}_{-0.0052}$ \\
        $\tau_\mathrm{reio}$         & $0.0592\pm0.0070$         & $0.0585^{+0.0065}_{-0.0076}$         & $0.0616\pm0.0073$         & $0.0595^{+0.0062}_{-0.0070}$ \\
        $\Omega_\mathrm{m}$          & $0.3008\pm0.0038$         & $0.2966\pm 0.0034$         & $0.3025\pm0.0036$         & $0.2978\pm 0.0032$           \\
        \hline
    \end{tabular}
\caption{The mean $\pm 1\sigma$ errors of cosmological parameters
for axion-like EDE fitting to the Planck and Planck+SPT+ACT
datasets with and without SH0ES. For upper limits, we quote the
$95\%$ confidence level.}
    \label{tab:axion}
\end{table}

\begin{figure}
    \centering
   \includegraphics[width=0.49\linewidth]{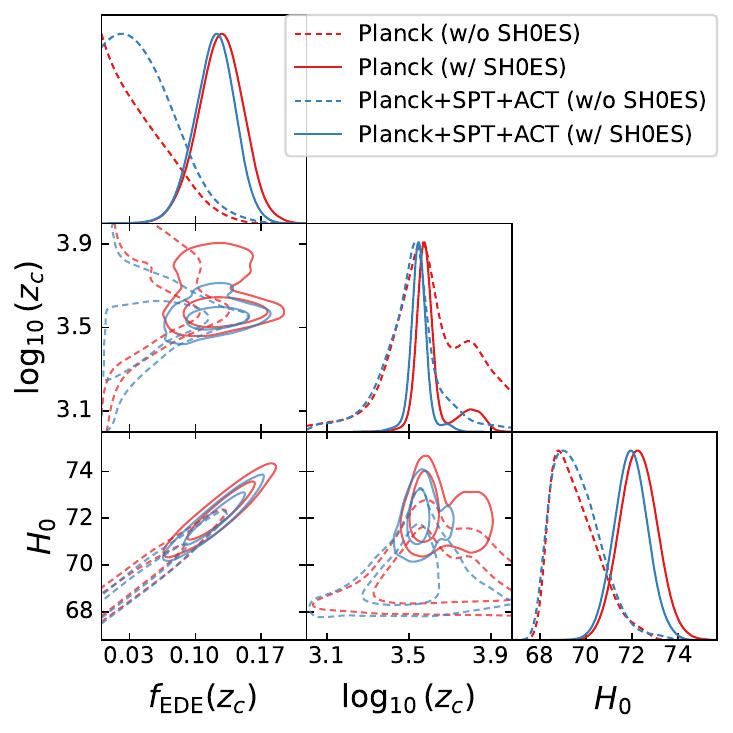}
    \hfill
   \includegraphics[width=0.49\linewidth]{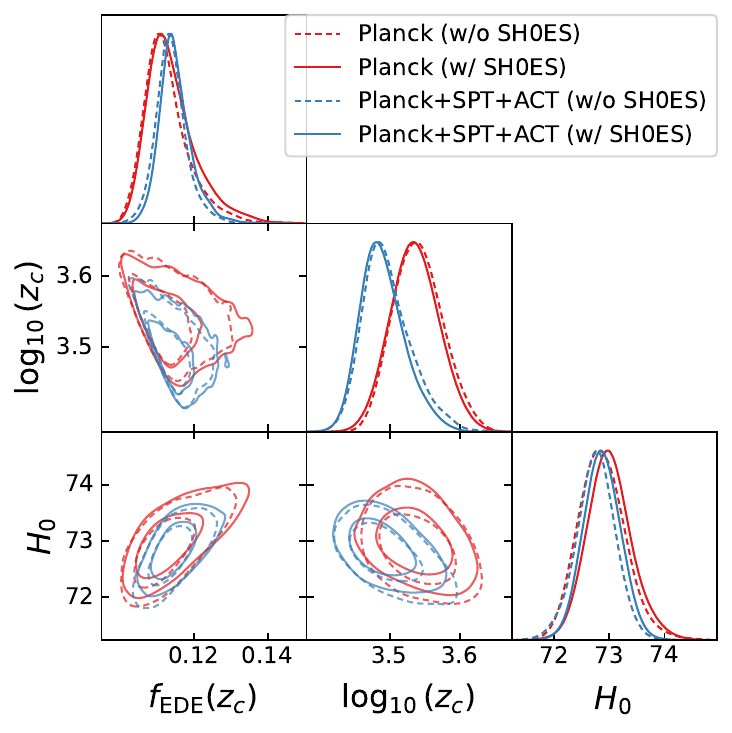}
\caption{1D and 2D marginalized posterior distributions ($68\%$
and $95\%$ confidence range) of relevant parameters for axion-like
EDE (left) and AdS-EDE (right), fitting to Planck and
Planck+SPT+ACT datasets with and without SH0ES. }
    \label{fig:results}
\end{figure}

%The latest ground-based CMB data leads to tighter constraints on
%axion-like EDE compared to previous ones. Unlike results with the
%previous ACT and SPT data which favored non-zero EDE fractions and
%large Hubble constants without a late-time $H_0$ prior
%\cite{Jiang:2022uyg,Smith:2022hwi,Peng:2023bik}, we can only
%observe $95\%$ upper limits on $f_\mathrm{EDE}(z_c)$ for
%axion-like EDE in both datasets without the SH0ES
%calibration.

The results for axion-like EDE without the SH0ES calibrated SN
dataset are $f_\mathrm{EDE}(z_c)<0.107$ and
$H_0=69.58^{+0.61}_{-1.3}$ km/s/Mpc for Planck, and
$f_\mathrm{EDE}(z_c)<0.105$ ($95\%$ upper limit on
$f_\mathrm{EDE}(z_c)$\footnote{Our result differs slightly from
that of Ref.~\cite{SPT-3G:2025vyw}, which reported a $68\%$ CL
lower limit for $f_\mathrm{EDE}(z_c)$ using similar datasets. We
%attribute this difference to the SN dataset we use and the
%$\tau_\mathrm{reio}$ prior they adopted,
clarify the origin of this difference in Appendix~\ref{sec:tau}.})
and $H_0=69.50^{+0.70}_{-1.2}$ km/s/Mpc for Planck+SPT+ACT. The
inclusion of ACT and SPT slightly tightens the constraints.
%Neither dataset shows a significant preference for axion-like EDE.
The results with the SH0ES calibration are $H_0\simeq 72$km/s/Mpc
for both datasets. In this case, $n_s=0.9921^{+0.0057}_{-0.0064}$
for Planck and $n_s=0.9897^{+0.0045}_{-0.0052}$ for
Planck+SPT+ACT, both are compatible with unity at the $2\sigma$
level.

The AdS-EDE model is known for yielding a larger $H_0$ even
without the SH0ES calibration, which here is seen again. The
results with Planck are
$f_\mathrm{EDE}(z_c)=0.1126^{+0.0037}_{-0.0071}$ and
$H_0=72.87^{+0.38}_{-0.45}$km/s/Mpc, while Planck+SPT+ACT leads to
slightly tighter constraints compared to Planck, with
$f_\mathrm{EDE}(z_c)=0.1137^{+0.0033}_{-0.0044}$ and $H_0=72.76\pm
0.356$ km/s/Mpc. The spectral index $n_s$ is highly consistent
with a scale-invariant HZ spectrum, $n_s=0.9975\pm 0.0043$ for
Planck and $n_s=0.9960\pm0.047$ for Planck+SPT+ACT. The results
with the SH0ES calibration are very similar, as shown in
Fig.~\ref{fig:results}.

\begin{figure}
    \centering
   \includegraphics[width=0.49\linewidth]{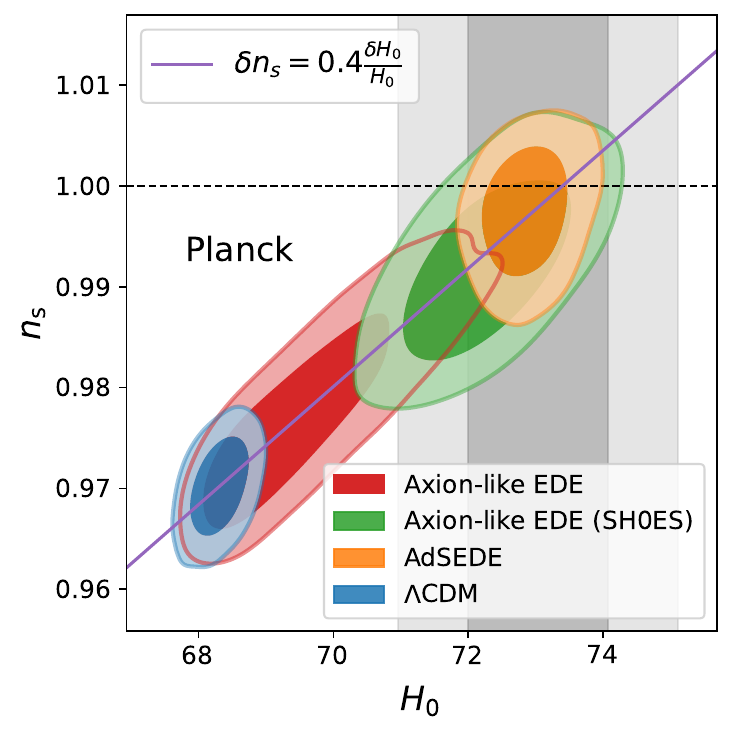}
    \hfill
   \includegraphics[width=0.49\linewidth]{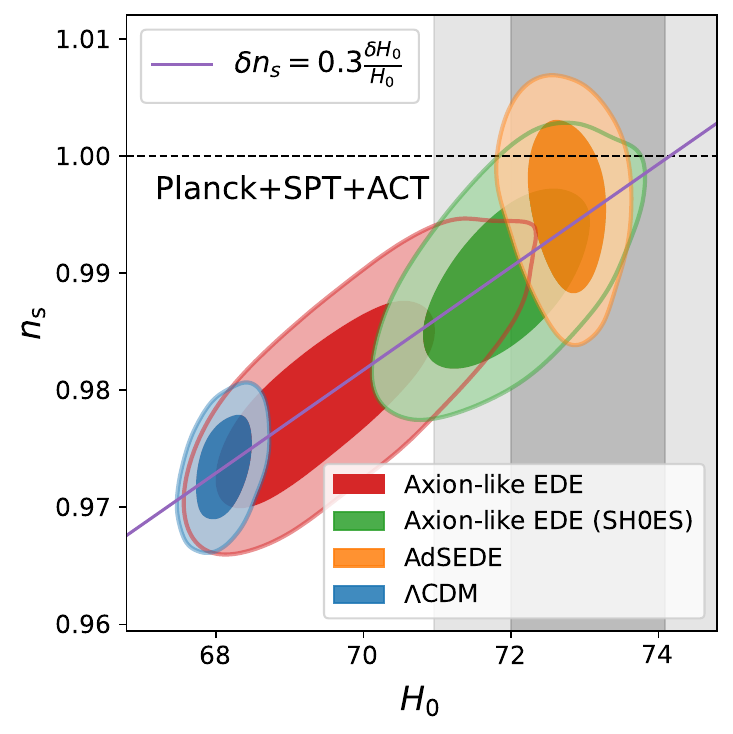}
\caption{The $n_s-H_0$ scaling relation from Planck (left) and
Planck+SPT+ACT (right). We present the $68\%$ and $95\%$ posterior
distributions for axion-like EDE with and without SH0ES, AdS-EDE
and $\Lambda$CDM. The purple lines are $\delta n_s=0.4\frac{\delta
H_0}{H_0}$ (left) and $\delta n_s=0.3\frac{\delta H_0}{H_0}$
(right) for Planck and Planck+SPT+ACT, respectively. The grey
bands are $1\sigma$ and $2\sigma$ regions of the latest $H_0$
measurement from SH0ES \cite{Riess:2021jrx}. }
    \label{fig:ns-H0}
\end{figure}

In Fig.~\ref{fig:ns-H0}, we present the $n_s-H_0$ scaling
relations for Planck and Planck+SPT+ACT datasets, respectively. As
seen, the scale relation (\ref{ACTSPTns}) is still robust.

%shows a clear $n_s-H_0$ scaling relation, except with a slightly
%smaller scale factor.

%\subsection{AdS-EDE}

%The mean and $1\sigma$ errors of cosmological parameters for
%AdS-EDE are presented in Table~\ref{tab:ads}. We also show the 1D
%and 2D marginalized posterior distributions of relevant parameters
%in the right panel of Fig.~\ref{fig:results}.

\begin{table}[htbp]
    \centering
    \begin{tabular}{|l|c|c|c|c|}
        \hline
        \multicolumn{1}{|l|}{Parameter} & \multicolumn{2}{c|}{Planck} & \multicolumn{2}{c|}{Planck+SPT+ACT} \\
        \cline{2-5}
        & w/o SH0ES & w/ SH0ES & w/o SH0ES & w/ SH0ES \\
        \hline
        $f_\mathrm{EDE}(z_c)$        & $0.1126^{+0.0037}_{-0.0071}$ & $0.1139^{+0.0040}_{-0.0079}$ & $0.1137^{+0.0034}_{-0.0044}$ & $0.1149^{+0.0032}_{-0.0048}$ \\
        $\log_{10}(z_c)$             & $3.541^{+0.032}_{-0.038}$    & $3.536\pm 0.035$             & $3.497^{+0.026}_{-0.040}$    & $3.491^{+0.025}_{-0.036}$    \\
        \hline
        $H_0$                        & $72.87^{+0.38}_{-0.45}$      & $73.01^{+0.36}_{-0.43}$      & $72.76\pm 0.36$              & $72.87\pm 0.34$              \\
        $100\Omega_\mathrm{b} h^2$   & $2.342\pm 0.018$             & $2.344^{+0.019}_{-0.016}$    & $2.306\pm0.014$    & $2.306\pm 0.014$             \\
        $\Omega_\mathrm{c} h^2$      & $0.1335\pm 0.0016$           & $0.1336\pm 0.0016$           & $0.1343^{+0.0013}_{-0.0011}$ & $0.1345\pm 0.0012$           \\
        $10^9A_\mathrm{s}$           & $2.167\pm 0.030$             & $2.169\pm 0.030$             & $2.141\pm 0.025$             & $2.142\pm 0.025$             \\
        $n_\mathrm{s}$               & $0.9975\pm 0.0043$           & $0.9977^{+0.0045}_{-0.0041}$           & $0.9960\pm 0.0047$ & $0.9957\pm 0.0048$           \\
        $\tau_\mathrm{reio}$         & $0.0546\pm 0.0072$           & $0.0547\pm 0.0072$           & $0.0483^{+0.0069}_{-0.0062}$           & $0.0483\pm 0.0068$           \\
        $\Omega_\mathrm{m}$          & $0.2967\pm 0.0037$           & $0.2959\pm 0.0035$           & $0.2986\pm 0.0033$           & $0.2980\pm 0.0033$           \\
        % \hline
        % $\chi^2$                  & $4203.50$                     & $12404.05$                    & $31994.57$                    \\
        \hline
    \end{tabular}
\caption{The mean $\pm 1\sigma$ errors of cosmological parameters
for AdS-EDE fitting to the Planck and Planck+SPT+ACT datasets with
and without SH0ES.}
    \label{tab:ads}
\end{table}

\section{Discussion}\label{sec:discussion}

In the EDE resolutions of the Hubble tension, the primordial
scalar spectral index must shift towards $n_s=1$ to compensate the
uplift of the bestfit value of $H_0$ so that $n_s=1$ for
$H_0\simeq 73$km/s/Mpc. In this work, we have tested this result
with latest ACT DR6 and SPT-3G D1 data, using two representative
EDE models, axion-like EDE and AdS-EDE.

It might be expected that high-precision small-scale SPT and ACT
data can be very powerful for constraining the spectral index
$n_s$ and EDE, which possibly disfavors the shift of $n_s$ towards
$n_s=1$. However, our results show that $n_s=1$ is not only
compatible with but in fact well-supported by the latest ACT and
SPT data.
%For axion-like EDE, when a high $H_0$ is preferred with the SH0ES
%calibration, the inferred $n_s$ is consistent with unity at the
%$2\sigma$ level. Furthermore, the AdS-EDE model yields a high
%$H_0$ even without the SH0ES calibration, showing high consistency
%with $n_s=1$ at the $1\sigma$ level.
The characteristic $n_s-H_0$ scaling relation for Planck+SPT+ACT,
i.e.,(\ref{ACTSPTns}), is still robust and is consistent with the
results using earlier ACT and SPT
\cite{Jiang:2022uyg,Smith:2022hwi,Peng:2023bik}. Therefore, the
prediction of $n_s=1$ in complete EDE resolution of the Hubble
tension is reconfirmed with the precise measurements from ACT and
SPT at high multipoles beyond the Planck angular resolution and
sensitivity.

%with the inclusion of the new ground-based CMB data. Specifically,
%as shown in the right panel of Fig.~\ref{fig:ns-H0}, the results
%from Planck+SPT+ACT shows a clear correlation between $n_s$ and
%$H_0$ but with a slightly smaller scale factor
%\begin{equation}
%    \delta n_s \simeq 0.3 \frac{\delta H_0}{H_0},
%\end{equation}

The fact that the resolution of the Hubble tension naturally leads
to $n_s=1$ has profound implications for our insight into
inflation and the primordial Universe, see e.g.
\cite{Kallosh:2022ggf,Ye:2022efx,Takahashi:2021bti,DAmico:2021fhz,Braglia:2022phb,Jiang:2023bsz,Giare:2023kiv,Huang:2023chx,Giare:2024akf}.
Our work again highlights the importance of re-examining our
understanding of the very early Universe within the broader
context of cosmological tensions.

\begin{acknowledgments}
This work
is supported by NSFC, No.12475064, National Key Research and
Development Program of China, No.2021YFC2203004, and the
Fundamental Research Funds for the Central Universities.
We acknowledge the use of high performance computing services provided by the International Centre for Theoretical Physics Asia-Pacific cluster.

\end{acknowledgments}

% \clearpage

\bibliographystyle{apsrev4-1}
\bibliography{ref}

\appendix

\section{The EDE models}\label{sec:ede}
In this Appendix, we briefly describe the EDE models used. In the
corresponding models, an unknown energy component, i.e.EDE,
behaves like a cosmological constant at $z\gtrsim 3000$ and then
decays rapidly before recombination, so that it suppresses the
sound horizon but does not affect the late evolution of the
Universe. The angular scale of sound horizon $r_s^*$ at
recombination is
\begin{equation}
    \theta_s^* = \frac{r_s^*}{D_A^*} \sim r_s^*H_0,
\end{equation}
which can be precisely set with CMB data, where $D_A^*$ is the
angular diameter distance to last scattering. Therefore,
%if the
%evolution after recombination follows flat $\Lambda$CDM,
we naturally have a higher value of $H_0$ for a lower $r_s^*$.

In this paper, we consider two well-known EDE models. The first is
axion-like EDE \cite{Poulin:2018cxd,Poulin:2018dzj}. In this
model, EDE is an ultra-light scalar field $\phi$ with an
axion-like potential:
\begin{equation}
V(\theta) = m^2f^2\left(1-\cos\theta\right)^n, \quad \theta \in
\left[-\pi,\pi\right]
\end{equation}
where $\theta \equiv \phi/f$ is the re-normalized field variable,
$m$ and $f$ are the effective mass and the couple constant of
axion-like EDE, respectively, see also
\cite{McDonough:2022pku,Cicoli:2023qri} for modelling it in string
theory. At early times, it is frozen at certain initial value,
$\theta_i=\phi_i/f$, due to the Hubble friction, and behaves like
dark energy. Afterwards, as the Hubble parameter falls, the field
will start to roll down at a critical redshift $z_c$ and rapidly
oscillate. As a result, the energy density of EDE will decay with
an equation of state $w\approx (n-1)/(n+1)$
\cite{PhysRevD.28.1243,Poulin:2018dzj}. In this work, we will set
$n=3$ following Ref.~\cite{Poulin:2018cxd}.

Another EDE model we consider is AdS-EDE \cite{Ye:2020btb}, in
which we have an AdS phase around recombination. In this work, we
consider a phenomenological potential\footnote{Other potentials
are also possible, see e.g. \cite{Ye:2020oix}.}:
\begin{equation}
    V(\phi)= \begin{dcases}
        V_0\left(\frac{\phi}{M_{\mathrm{Pl}}}\right)^4-V_{\mathrm{AdS}}, & \frac{\phi}{M_{\mathrm{Pl}}}<\left(\frac{V_{\mathrm{AdS}}}{V_0}\right)^{1 / 4} \\ 0 , & \frac{\phi}{M_{\mathrm{Pl}}}>\left(\frac{V_{\mathrm{AdS}}}{V_0}\right)^{1 / 4}
    \end{dcases}
\end{equation}
where $V_{\mathrm{AdS}}$ is the depth of the AdS well,
$M_{\mathrm{Pl}}$ is the reduced Planck mass. The implications of
AdS vacuum for our current Universe and inflation in early
Universe also have been studied in recent
Refs. \cite{Visinelli:2019qqu,Akarsu:2019hmw,Akarsu:2021fol,Akarsu:2022typ,Sen:2021wld,DiGennaro:2022ykp,Ong:2022wrs,Malekjani:2023ple,Adil:2023exv,Adil:2023ara,Wang:2024hwd,Wang:2025dtk}
and
e.g. Ref. \cite{Felder:2002jk,Piao:2004hr,Piao:2005ag,Li:2019ipk,Li:2020cjj},
respectively. The existence of an AdS phase makes the energy
density of EDE decay faster than in oscillation phase. Therefore,
compared to axion-like EDE, AdS-EDE can allow a more efficient
injection of EDE with less influence on the fit to CMB data. As a
result, AdS-EDE has the advantage of yielding a large Hubble
constant, $H_0\simeq 73$ km/s/Mpc, without the inclusion of any
$H_0$ prior \cite{Ye:2020btb,Ye:2020oix,Jiang:2021bab,Jiang:2022uyg,Wang:2024tjd}.

\section{The effects of $\tau_{\mathrm{reio}}$ prior and SN data} \label{sec:tau}

The recent Ref.~\cite{SPT-3G:2025vyw} reported
$f_\mathrm{EDE}(z_c)=0.071^{+0.035}_{-0.038}$ at $68\%$ CL for
axion-like EDE when using the combined Planck+SPT+ACT dataset and
DESI BAO, without the SH0ES calibration.  This result is in mild
tension with ours using similar datasets, which only shows a
$95\%$ upper limits on $f_\mathrm{EDE}(z_c)$. We attribute this
difference mainly to the $\tau_{\mathrm{reio}}$ prior they adopted
and the SN dataset we include.

\begin{figure}[htbp]
    \centering
   \includegraphics[width=\linewidth]{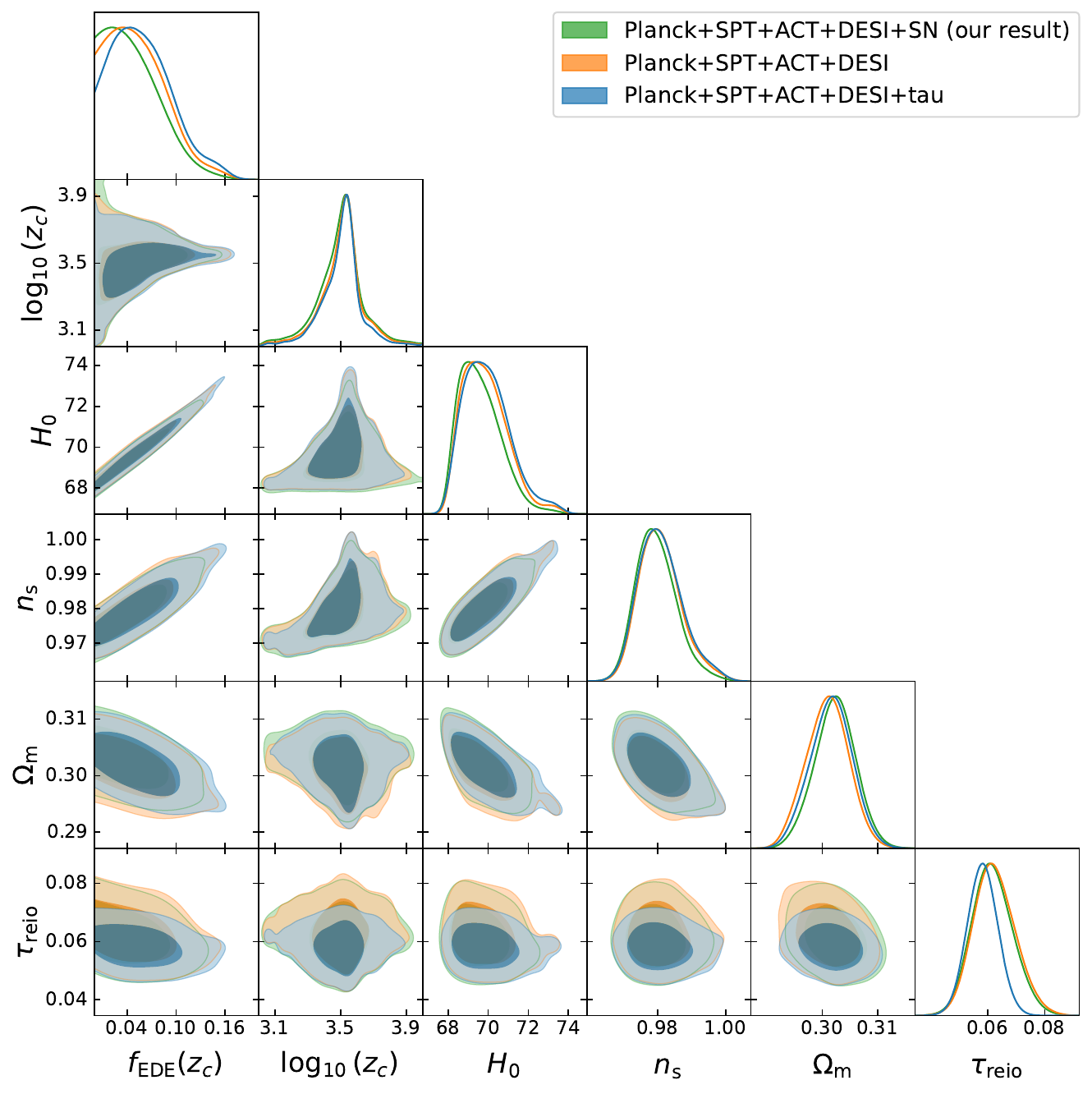}
\caption{1D and 2D marginalized posterior distributions ($68\%$
and $95\%$ confidence range) of relevant parameters for axion-like
EDE. The plot shows our baseline results against a reweighted
analysis that excludes the SN data and adopts the
$\tau_{\mathrm{reio}}$ prior from Ref.~\cite{SPT-3G:2025vyw}. }
    \label{fig:tau}
\end{figure}

Ref.~\cite{SPT-3G:2025vyw} adopted a Gaussian prior on the optical
depth of reionization, i.e. $\tau_{\mathrm{reio}}=0.051\pm 0.006$,
in place of the Planck low-$\ell$ EE likelihood, and did not
include the Pantheon+ SN data. To clarify the origin of the
difference, we perform a reweighting of our MCMC chains using the
post-process of \texttt{cobaya}, removing the SN data we use and
also further adopting the same $\tau_{\mathrm{reio}}$ prior as in
Ref.~\cite{SPT-3G:2025vyw}.

As shown in Fig.~\ref{fig:tau}, removing the SN data slightly
relaxes the constraints on $f_\mathrm{EDE}(z_c)$, possibly due to
the influence of the Pantheon+ data on $\Omega_\mathrm{m}$.
Importantly, when we further replace the Planck low-$\ell$ EE
likelihood with the $\tau_{\mathrm{reio}}$ prior, i.e.
Planck+SPT+ACT+DESI+tau, we also observe a $68\%$ CL lower limit
on the EDE fraction,
$f_\mathrm{EDE}(z_c)=0.060^{+0.024}_{-0.049}$, consistent with the
results in Ref.~\cite{SPT-3G:2025vyw} using the same dataset. This
indicates that the results in Ref.~\cite{SPT-3G:2025vyw} are
caused by the specific manipulation for $\tau_{\mathrm{reio}}$.

\end{document}